\documentclass[%
reprint,
superscriptaddress,
 amsmath,amssymb,
]{revtex4-2}
\usepackage{float}
\usepackage{xcolor}
\usepackage{graphicx}
\usepackage{dcolumn}
\usepackage{bm}

\begin{document}

\title{Measurement of the Saturation Length of the Self-Modulation Instability}

\author{A.~Clairembaud}
\affiliation{Max Planck Institute for Physics, 85748 Garching bei München, Germany}

\author{M.~Turner}
\affiliation{CERN, 1211 Geneva 23, Switzerland}

\author{M.~Bergamaschi}
\affiliation{CERN, 1211 Geneva 23, Switzerland}

\author{L.~Ranc}
\affiliation{Max Planck Institute for Physics, 85748 Garching bei München, Germany}

\author{F.~Pannell}
\affiliation{UCL, London WC1E 6BT, United Kingdom}

\author{J.~Mezger}
\affiliation{Max Planck Institute for Physics, 85748 Garching bei München, Germany}

\author{H.~Jaworska}
\affiliation{CERN, 1211 Geneva 23, Switzerland}

\author{N.~van Gils}
\affiliation{CERN, 1211 Geneva 23, Switzerland}
\affiliation{PARTREC, UMCG, University of Groningen, Groningen, NL}

\author{J.~Farmer}
\affiliation{Max Planck Institute for Physics, 85748 Garching bei München, Germany}

\author{P.~Muggli}
\affiliation{Max Planck Institute for Physics, 85748 Garching bei München, Germany}
\affiliation{CERN, 1211 Geneva 23, Switzerland}

\collaboration{AWAKE Collaboration}

\author{C.C.~Ahdida}
\affiliation{CERN, 1211 Geneva 23, Switzerland}
\author{Y. Alekajbaf}
\affiliation{Uppsala University, Uppsala, Sweden}
\author{C.~Amoedo}
\affiliation{CERN, 1211 Geneva 23, Switzerland}
\author{O.~Apsimon}
\affiliation{University of Manchester, Manchester M13 9PL, United Kingdom}
\affiliation{Cockcroft Institute, Warrington WA4 4AD, United Kingdom}
\author{R.~Apsimon}
\affiliation{Cockcroft Institute, Warrington WA4 4AD, United Kingdom}
\affiliation{Lancaster University, Lancaster LA1 4YB, United Kingdom}
\author{T.~Bachmann}
\affiliation{CERN, 1211 Geneva 23, Switzerland}
\author{C.~Badiali}
\affiliation{GoLP/Instituto de Plasmas e Fus\~{a}o Nuclear, Instituto Superior T\'{e}cnico, Universidade de Lisboa, 1049-001 Lisbon, Portugal}
\author{M.~Baquero}
\affiliation{Ecole Polytechnique Federale de Lausanne (EPFL), Swiss Plasma Center (SPC), 1015 Lausanne, Switzerland}
\author{E.~Belli}
\affiliation{CERN, 1211 Geneva 23, Switzerland}
\affiliation{John Adams Institute, Oxford University, Oxford OX1 3RH, United Kingdom}
\author{A.~Boccardi}
\affiliation{CERN, 1211 Geneva 23, Switzerland}
\author{T.~Bogey}
\affiliation{CERN, 1211 Geneva 23, Switzerland}
\author{S.~Burger}
\affiliation{CERN, 1211 Geneva 23, Switzerland}
\author{P.N.~Burrows}
\affiliation{John Adams Institute, Oxford University, Oxford OX1 3RH, United Kingdom}
\author{B.~Buttensch{\"o}n}
\affiliation{Max Planck Institute for Plasma Physics, 17491 Greifswald, Germany}
\author{A.~Caldwell}
\affiliation{Max Planck Institute for Physics, 85748 Garching bei München, Germany}
\author{M.~Chung}
\affiliation{POSTECH, Pohang 37673, Republic of Korea}
\author{C.C.~Cobo}
\affiliation{Imperial College London, London SW7 2AZ, United Kingdom}
\author{D.A.~Cooke}
\affiliation{UCL, London WC1E 6BT, United Kingdom}
\author{D.~Dancila}
\affiliation{Uppsala University, Uppsala, Sweden}
\author{C.~Davut}
\affiliation{University of Manchester, Manchester M13 9PL, United Kingdom}
\affiliation{Cockcroft Institute, Warrington WA4 4AD, United Kingdom}
\author{G.~Demeter}
\affiliation{HUN-REN Wigner Research Centre for Physics, Budapest, Hungary}
\author{A.C.~Dexter}
\affiliation{Cockcroft Institute, Warrington WA4 4AD, United Kingdom}
\affiliation{Lancaster University, Lancaster LA1 4YB, United Kingdom}
\author{S.~Doebert}
\affiliation{CERN, 1211 Geneva 23, Switzerland}
\author{A.~Eager}
\affiliation{CERN, 1211 Geneva 23, Switzerland}
\author{D.~Easton}
\affiliation{GWA, Cambridge, CB4 0WS UK}
\author{B.~Elward}
\affiliation{University of Wisconsin, Madison, WI 53706, USA}
\author{R.~Fonseca}
\affiliation{DCTI/ISCTE, Instituto Universit\'{e}rio de Lisboa, 1649-026, Lisboa, Portugal}
\affiliation{GoLP/Instituto de Plasmas e Fus\~{a}o Nuclear, Instituto Superior T\'{e}cnico, Universidade de Lisboa, 1049-001 Lisbon, Portugal}
\author{I.~Furno}
\affiliation{Ecole Polytechnique Federale de Lausanne (EPFL), Swiss Plasma Center (SPC), 1015 Lausanne, Switzerland}
\author{K.~Gajewski}
\affiliation{Uppsala University, Uppsala, Sweden}
\author{D.~Ghosal}
\affiliation{Cockcroft Institute, Warrington WA4 4AD, United Kingdom}
\affiliation{University of Liverpool, Liverpool L69 7ZE, United Kingdom}
\author{E.~Granados}
\affiliation{CERN, 1211 Geneva 23, Switzerland}
\author{J.~Gregory}
\affiliation{Cockcroft Institute, Warrington WA4 4AD, United Kingdom}
\affiliation{Lancaster University, Lancaster LA1 4YB, United Kingdom}
\author{O.~Grulke}
\affiliation{Max Planck Institute for Plasma Physics, 17491 Greifswald, Germany}
\affiliation{Technical University of Denmark, 2800 Kgs. Lyngby, Denmark}
\author{E.~Gschwendtner}
\affiliation{CERN, 1211 Geneva 23, Switzerland}
\author{E.~Guran}
\affiliation{CERN, 1211 Geneva 23, Switzerland}
\author{D.~Harryman}
\affiliation{CERN, 1211 Geneva 23, Switzerland}
\author{M.~Hibberd}
\affiliation{University of Manchester, Manchester M13 9PL, United Kingdom}
\affiliation{Cockcroft Institute, Warrington WA4 4AD, United Kingdom}
\author{P.~Karataev}
\affiliation{John Adams Institute, Oxford University, Oxford OX1 3RH, United Kingdom}
\affiliation{Royal Holloway University of London, Egham, Surrey, TW20 0EX, United Kingdom}
\author{R.~Karimov}
\affiliation{Ecole Polytechnique Federale de Lausanne (EPFL), Swiss Plasma Center (SPC), 1015 Lausanne, Switzerland}
\author{M.A.~Kedves}
\affiliation{HUN-REN Wigner Research Centre for Physics, Budapest, Hungary}
\author{F.~Kraus}
\affiliation{Universität Bonn, 53121 Bonn, Germany}
\author{M.~Krupa}
\affiliation{CERN, 1211 Geneva 23, Switzerland}
\author{T.~Lefevre}
\affiliation{CERN, 1211 Geneva 23, Switzerland}
\author{T.~Lofnes}
\affiliation{Uppsala University, Uppsala, Sweden}
\author{N.~Lopes}
\affiliation{GoLP/Instituto de Plasmas e Fus\~{a}o Nuclear, Instituto Superior T\'{e}cnico, Universidade de Lisboa, 1049-001 Lisbon, Portugal}
\author{K.~Lotov}
\noaffiliation
\author{J.~Mcgunigal}
\affiliation{University of Manchester, Manchester M13 9PL, United Kingdom}
\affiliation{Cockcroft Institute, Warrington WA4 4AD, United Kingdom}
\author{S.A.~Mohadeskasaei}
\affiliation{Uppsala University, Uppsala, Sweden}
\author{M.~Moreira}
\affiliation{CERN, 1211 Geneva 23, Switzerland}
\author{B.~Moser}
\affiliation{CERN, 1211 Geneva 23, Switzerland}
\author{Z.~Najmudin}
\affiliation{Imperial College London, London SW7 2AZ, United Kingdom}
\author{S.~Norman}
\affiliation{University of Manchester, Manchester M13 9PL, United Kingdom}
\affiliation{Cockcroft Institute, Warrington WA4 4AD, United Kingdom}
\author{N.~Okhotnikov}
\noaffiliation
\author{A.~Omoumi}
\affiliation{CERN, 1211 Geneva 23, Switzerland}
\author{C.~Pakuza}
\affiliation{CERN, 1211 Geneva 23, Switzerland}
\author{A.~Pardons}
\affiliation{CERN, 1211 Geneva 23, Switzerland}
\author{K.~Pelckmans}
\affiliation{Uppsala University, Uppsala, Sweden}
\author{J.~Pisani}
\affiliation{GWA, Cambridge, CB4 0WS UK}
\author{A.~Pukhov}
\affiliation{Heinrich-Heine-Universit{\"a}t D{\"u}sseldorf, 40225 D{\"u}sseldorf, Germany}
\author{R.~Rossel}
\affiliation{CERN, 1211 Geneva 23, Switzerland}
\author{H.~Saberi}
\affiliation{University of Manchester, Manchester M13 9PL, United Kingdom}
\affiliation{Cockcroft Institute, Warrington WA4 4AD, United Kingdom}
\author{M.V.~dos Santos}
\affiliation{CERN, 1211 Geneva 23, Switzerland}
\author{O.~Schmitz}
\affiliation{University of Wisconsin, Madison, WI 53706, USA}
\author{F.~Sharmin}
\affiliation{University of Wisconsin, Madison, WI 53706, USA}
\author{F.~Silva}
\affiliation{INESC-ID, Instituto Superior Técnico, Universidade de Lisboa, 1049-001 Lisbon, Portugal}
\author{L.~Silva}
\affiliation{GoLP/Instituto de Plasmas e Fus\~{a}o Nuclear, Instituto Superior T\'{e}cnico, Universidade de Lisboa, 1049-001 Lisbon, Portugal}
\author{B.~Spear}
\affiliation{John Adams Institute, Oxford University, Oxford OX1 3RH, United Kingdom}
\author{L.~Stant}
\affiliation{CERN, 1211 Geneva 23, Switzerland}
\author{C.~Stollberg}
\affiliation{Ecole Polytechnique Federale de Lausanne (EPFL), Swiss Plasma Center (SPC), 1015 Lausanne, Switzerland}
\author{A.~Sublet}
\affiliation{CERN, 1211 Geneva 23, Switzerland}
\author{C.~Swain}
\affiliation{Cockcroft Institute, Warrington WA4 4AD, United Kingdom}
\affiliation{University of Liverpool, Liverpool L69 7ZE, United Kingdom}
\author{G.~Tenasini}
\affiliation{CERN, 1211 Geneva 23, Switzerland}
\author{A.~Topaloudis}
\affiliation{CERN, 1211 Geneva 23, Switzerland}
\author{P.~Tuev}
\noaffiliation
\author{J.~Uncles}
\affiliation{GWA, Cambridge, CB4 0WS UK}
\author{F.~Velotti}
\affiliation{CERN, 1211 Geneva 23, Switzerland}
\author{J.~Vieira}
\affiliation{GoLP/Instituto de Plasmas e Fus\~{a}o Nuclear, Instituto Superior T\'{e}cnico, Universidade de Lisboa, 1049-001 Lisbon, Portugal}
\author{C.~Welsch}
\affiliation{Cockcroft Institute, Warrington WA4 4AD, United Kingdom}
\affiliation{University of Liverpool, Liverpool L69 7ZE, United Kingdom}
\author{T.~Wilson}
\affiliation{Heinrich-Heine-Universit{\"a}t D{\"u}sseldorf, 40225 D{\"u}sseldorf, Germany}
\author{M.~Wing}
\affiliation{UCL, London WC1E 6BT, United Kingdom}
\author{J.~Wolfenden}
\affiliation{Cockcroft Institute, Warrington WA4 4AD, United Kingdom}
\affiliation{University of Liverpool, Liverpool L69 7ZE, United Kingdom}
\author{B.~Woolley}
\affiliation{CERN, 1211 Geneva 23, Switzerland}
\author{G.~Xia}
\affiliation{Cockcroft Institute, Warrington WA4 4AD, United Kingdom}
\affiliation{University of Manchester, Manchester M13 9PL, United Kingdom}
\author{V.~Yarygova}
\noaffiliation
\author{W.~Zhang}
\affiliation{John Adams Institute, Oxford University, Oxford OX1 3RH, United Kingdom}

\date{\today}

\begin{abstract}
The self-modulation (SM) instability transforms a long charged particle bunch traveling in plasma into a train of microbunches that resonantly drives large-amplitude wakefields. We present the first determination of the saturation length of SM using experimental and numerical results. The saturation length is the distance over which wakefields reach their maximum amplitude along the plasma. By varying the plasma length and measuring the radius of the transverse distribution of the bunch, we find that the saturation length of SM decreases with plasma density and initial field amplitude, e.g., when seeding. The saturation length is a fundamental parameter of the instability, and these results are key for understanding SM and designing plasma wakefield accelerators driven by long bunches, such as AWAKE, or by long laser pulses for radiation production.

\end{abstract}

\maketitle

\clearpage
\emph{Introduction} -- Instabilities play a central role in plasma physics (see, e.g., Ref.~\cite{Hasegawa1975}). %
In particular, understanding their development and saturation is essential. %
A fundamental parameter characterizing many instabilities is the saturation length. %
An instability can grow in space from an initially small (linear) amplitude. %
Growth leads to nonlinear effects, and at the saturation length the amplitude reaches a large and essentially constant value, limited by these effects. %
In general, the effect sought after is maximum at saturation, and its event-to-event variations are also smaller than during growth. %

For example, in free electron lasers (FELs), a particle bunch modulates to emit light with exponentially growing intensity.
The saturation length, $L_\mathrm{sat}$, can be determined by measuring the radiation power as a function of undulator length (see e.g., Ref.~\cite{LCLSgrowth}). %
Determining $L_\mathrm{sat}$ is crucial, as it sets the minimum undulator length required for the FEL to reach maximum output power. %

In plasma, an analogous microbunching process occurs in the self-modulation (SM) instability of a relativistic charged particle bunch. %
SM develops when the duration of the bunch $\sigma_t$ is much longer than the plasma electron period $\tau_{pe}$~\cite{kumar}. %
It has been observed with electron~\cite{fang,DESYSM,frascatiSM} and proton~\cite{karl} bunches. %
SM also occurs with long laser pulses propagating in plasmas~\cite{SMLWFAtheo1,SMLWFAtheo3,SMLWFAtheo4} and has been observed in experiments~\cite{SMLWFAexp}. %
Despite these observations, the saturation length of SM has not yet been measured experimentally. %

The initial transverse wakefields driven by the long bunch have small amplitude and periodically modulate the bunch density with period $\tau_{pe}$. %
Since the modulated bunch drives higher-amplitude wakefields, a feedback loop between density modulation and wakefields is set up.
This leads to growth of SM along the plasma and along the bunch. %
The growth saturates when the bunch is fully modulated, i.e., when the resulting train of microbunches resonantly excites wakefields to the highest amplitudes. %

Unlike in FELs, where radiation power can be measured directly, wakefield amplitudes are difficult to measure~\cite{wan2024real}.
It is therefore necessary to rely on indirect effects of the wakefields to determine the saturation length of SM. %

Because SM is a transverse process, the formation of the microbunch train leads to that of a halo of defocused particles surrounding the train. %
This halo formation is both the cause (modulation of the bunch density) and a consequence (effect of transverse wakefields) of the growth of the wakefields along the plasma~\cite{marlenePRL}. %
One can expect the formation process to slow significantly once the microbunch train is fully formed, and when saturation of SM is reached, at $L_\mathrm{sat}$. %
This key parameter can therefore be experimentally determined by characterizing the evolution of the halo of defocused particles along the plasma~\cite{NIMAarthur}. %

In this \emph{Letter}, we determine for the first time, with experimental and numerical simulation results, the saturation length of the SM process. %
By varying the plasma length and measuring the radius of the halo at a screen downstream of the plasma, we identify the distance over which this radius increases and saturates as a function of experimental parameters. %
We find that the saturation length of the radius decreases with increasing plasma electron density, and that seeding the SM process makes this length shorter than in the non-seeded case. %
Numerical simulation results indicate that the saturation length of the halo radius provides a good estimate for that of SM. %
There is excellent agreement between numerical simulation and experimental results for the saturation length of the halo radius. %

\begin{figure}[b!]
    \centering
    \includegraphics[width=1\linewidth]{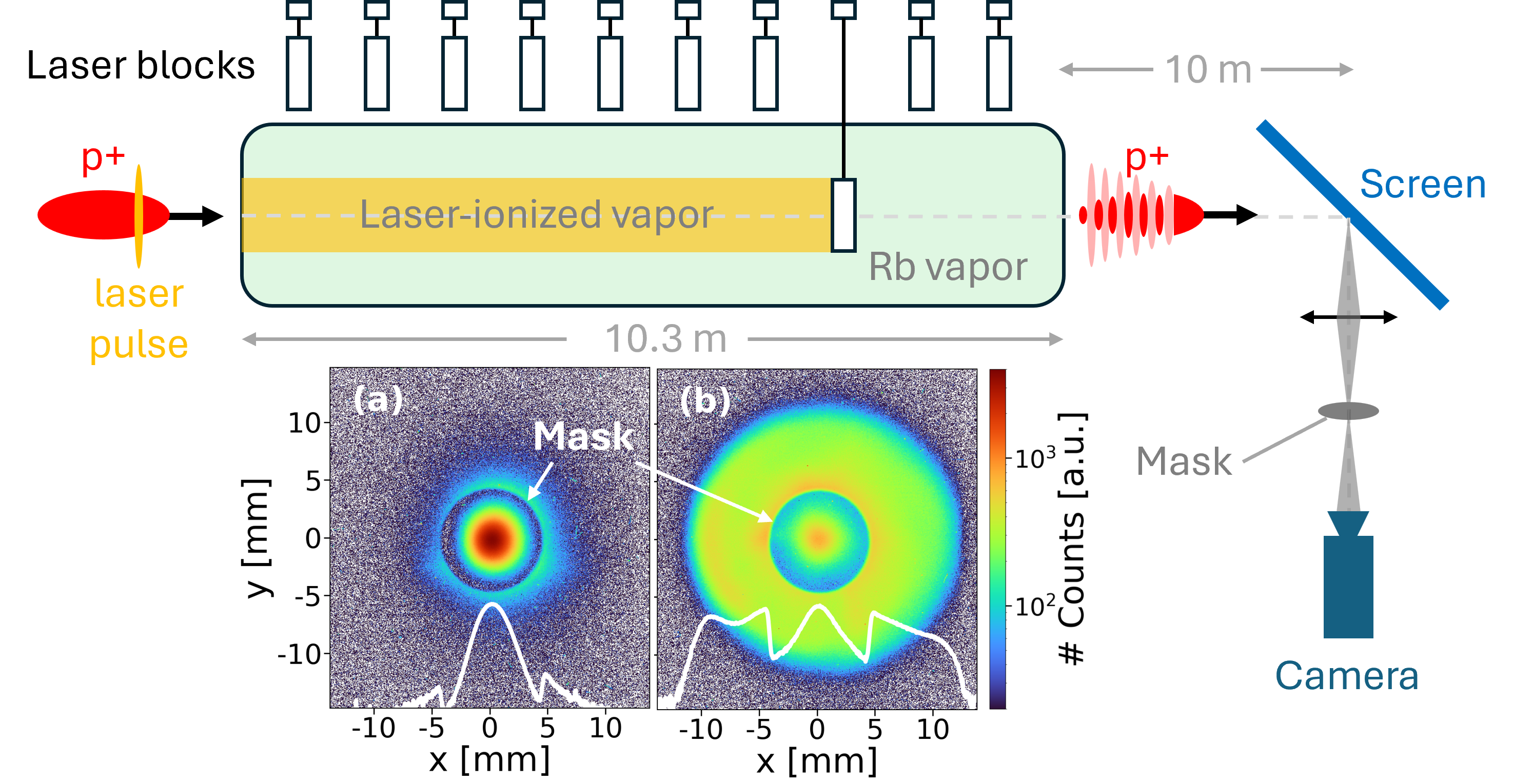}
    \caption{Schematic of the experimental setup (not to scale). %
    Inset: Transverse distribution of the bunch at the screen for single events after propagating through (a) Rubidium vapor, (b) $9.5\,$m of plasma. Colormap: logarithmic scale.}
    \label{fig:experimental_setup}
\end{figure}

\emph{Experimental Setup} -- Measurements were performed in the context of AWAKE (Advanced WAKefield Experiment) at CERN~\cite{AWAKESymmetry}. %
A schematic layout of the experiment is shown on Fig.~\ref{fig:experimental_setup}. %

\begin{figure*}
     \centering
     \includegraphics[width=1.0\linewidth]{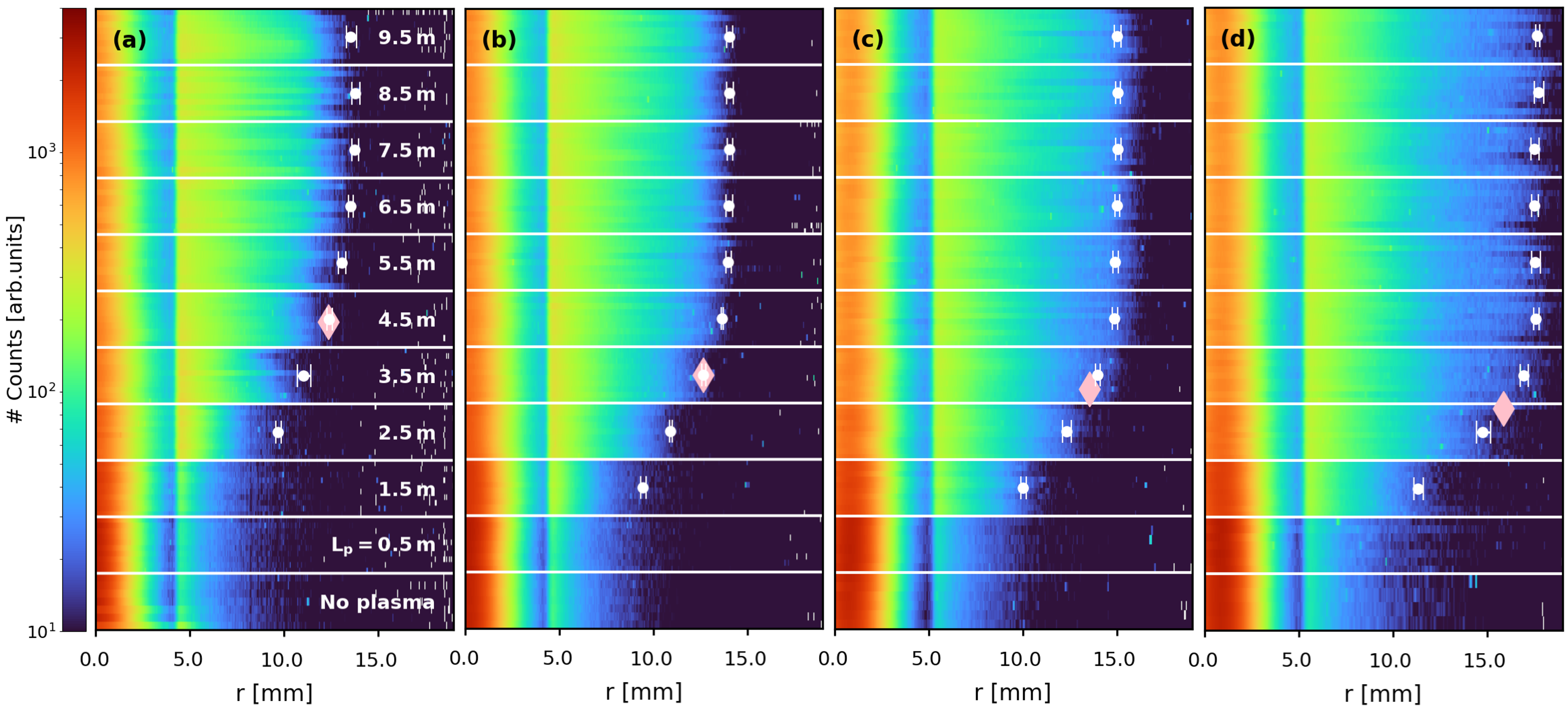}
     \caption{Radial slices of the transverse distribution of the bunch at the screen for different propagation distances in plasma ($L_p$). %
     (a) $n_{pe}=(1.06 \pm 0.01)\times10^{14}\,$cm$^{-3}$, 
     (b) $(1.98\pm 0.01)\times10^{14}\,$cm$^{-3}$, 
     (c) $(3.90\pm 0.02)\times10^{14}\,$cm$^{-3}$, 
     (d) $(7.42\pm 0.03)\times10^{14}\,$cm$^{-3}$.
     Colormap: logarithmic scale. %
     White symbols: average radius of the halo $r_h$ for the corresponding $L_p$, error bars: standard deviation of $\sim10$ events. %
     Pink diamonds: saturation point of the radius ($r_h = 0.9 \,r_{h,\mathrm{max}}$). %
     $t_\mathrm{RIF} = 100\,$ps ($\sim0.6\sigma_t$). 
     (a) $N_b =(2.98 \pm 0.08)\times10^{11}$,
     (b) $(2.98 \pm 0.03)\times10^{11}$, 
     (c) $(2.86 \pm 0.05)\times10^{11}$, 
     (d) $(2.87 \pm 0.05)\times10^{11}$.
     }
     \label{fig:waterfall}
 \end{figure*}
 
A $400\,$GeV proton bunch from the Super Proton Synchrotron, with a population $N_b \cong 3\times10^{11}$ particles, is focused to an rms transverse size $\sigma_r \cong 200\,$\textmu m at the entrance of the plasma. %
The bunch has a normalized emittance $\epsilon_N \cong 2.2\,$mm-mrad, and an rms duration $\sigma_t \cong 170\,$ps. %

A short laser pulse with FWHM duration 120\,fs and energy $\sim$100\,mJ co-propagates with the proton bunch. %
It forms a relativistic ionization front (RIF), and ionizes a rubidium vapor (RbI $\rightarrow$ RbII), resulting in a plasma electron density $n_{pe}$ that can be varied in the range $\cong (1 - 10)\times10^{14}\,$cm$^{-3}$~\cite{plyushchev2017rubidium}. %
This range of densities corresponds to plasma periods much shorter than the bunch duration (
$\tau_{pe} = 2\pi\sqrt{\frac{m_e \epsilon_0}{n_{pe}e^2} } \cong (11- 3)$\,ps $\ll\sigma_t \cong 170\,$ps), enabling the development of SM. %
When placed within the bunch, the RIF defines the onset of the beam-plasma interaction, and thus the timing and amplitude of the initial wakefields that can seed SM~\cite{fabian}. %

The laser pulse propagation can be stopped at different locations along the $10.3\,$m-long vapor column by inserting a thin ($200\,$ \textmu m) aluminum foil (laser blocks on Fig.~\ref{fig:experimental_setup}). %
The plasma length ($L_p$) over which the proton bunch propagates can therefore be varied ($L_p=0.5, 1.5, ..., 9.5\,$m), and thus also the distance over which SM develops. %

To observe the development of SM along the plasma, we record the transverse profile of the bunch on a scintillating screen as a function of $L_p$. %
The screen is placed $20.3\,$m downstream from the plasma entrance, and a camera images the light emitted by the screen after the passage of the bunch. %
Since the energy spread of the proton bunch caused by interaction with the wakefields is small ($\Delta E/E < 2.5\%$), %
the number of counts per pixel recorded by the camera is proportional to the local, time-integrated proton bunch density. %
A circular attenuating mask with transmission $17\%$ and radius $4\,$mm is placed in an image plane of the optical line of the camera, and approximately aligned with the center of the bunch (Inset (a), Fig.~\ref{fig:experimental_setup}). %
This enables simultaneous observation of the faint halo of defocused particles and of the intense bunch core formed by the train, without saturating the 12-bit camera. %

\emph{Results} -- The inset of Fig.~\ref{fig:experimental_setup} shows time-integrated transverse profiles of the bunch at the screen after propagation (a) in vapor (no laser pulse), and (b) in a plasma with $L_p=9.5\,$m. %
When propagating in vapor, the unmodulated bunch fits entirely under the mask and has a Gaussian-like distribution (Inset~(a)). %
After propagating in plasma, the bunch exhibits a halo, indicating that SM has developed~\cite{marlenePRL} (Inset~(b)). %
The halo distribution has a sharp outer edge, the radius of which, $r_h$, can be accurately determined following the procedure described in Ref.~\cite{turner2018method}. %

We extract a horizontal slice of the distribution of $\sim 10$ successive events for each value of $L_p$, and plot them consecutively on Fig.~\ref{fig:waterfall} for four values of $n_{pe}$. %
The instability is seeded, thus ensuring that the process is reproducible from event to event, as shown by the small variations of $r_h$.
Under these conditions, varying $L_p$ is equivalent to observing the state of the bunch (and related wakefields) at different locations along the plasma for a single SM event. %

The evolution of the bunch distribution (discussed below for Fig.~\ref{fig:waterfall}(d)) can be divided into three distinct phases. %

During the first phase, from $L_p = 0$ to $0.5\,$m, the seed fields driven by the unmodulated bunch have small amplitude (few $\text{MV/m}$)~\cite{fabian,marlenePRL}, and are predominantly focusing~\cite{keinigs1987two,liviofocusing}. %
Since the relativistic protons ($\gamma_0=427$) have a large inertia, very little transverse evolution occurs over this short distance in plasma (compare with ``No plasma''). %

During the second phase, occurring from $L_p = 0.5$ to $3.5\,$m, the periodic, transverse wakefields develop alternating focusing and defocusing regions. %
As a result, a fraction of the protons are defocused and leave the wakefields transversely, thus forming the halo surrounding the bunch core. %
This is seen %
on the image as an increase in charge density at large radii ($r>4\,$mm, i.e., beyond the mask) and as a simultaneous decrease under the mask, and thus to an increase in $r_h$, as expected. %
The observation of the formation of the halo with $L_p$ clearly indicates that SM is growing within this section of plasma. %
We quantify the evolution by measuring the average radius $r_{h,\mathrm{avg}}$ (white symbols), e.g., with an increase from $r_{h,\mathrm{avg}} = 11.3$ to $16.9\,$mm for $L_p = 1.5$ to $3.5\,$m. %

During the third phase, occurring from $L_p = 3.5$ to $9.5\,$m, the transverse distribution of the bunch remains essentially unchanged, and $r_h$ has saturated at its maximum value $r_{h,\mathrm{max}}$.
This is the signature expected from saturation of the SM process. %
We note that for these $L_p$, time-resolved images of the bunch (not shown) exhibit deep modulation of the longitudinal bunch density, another signature of the saturation of SM \cite{kumar,schroeder}. %

We define the saturation length of the halo radius $L_\mathrm{sat}$ as the interpolated $L_p$ at which $r_h$ reaches $0.9\, r_{h,\mathrm{max}}$: a threshold chosen to enable reliable determination of $L_\mathrm{sat}$, i.e., before the growth curve flattens and $r_h$ becomes weakly dependent on $L_p$. %
It is indicated by the pink diamonds on Fig.~\ref{fig:waterfall}. %
These clearly indicate the decrease of $L_\mathrm{sat}$ with increasing $n_{pe}$ (($4.5\pm0.2)$$\,$m, to ($2.9\pm0.1)$$\,$m from Figs.~\ref{fig:waterfall}(a) to (d)). %

\begin{figure}[b!]
    \centering
    \includegraphics[width=1\linewidth]{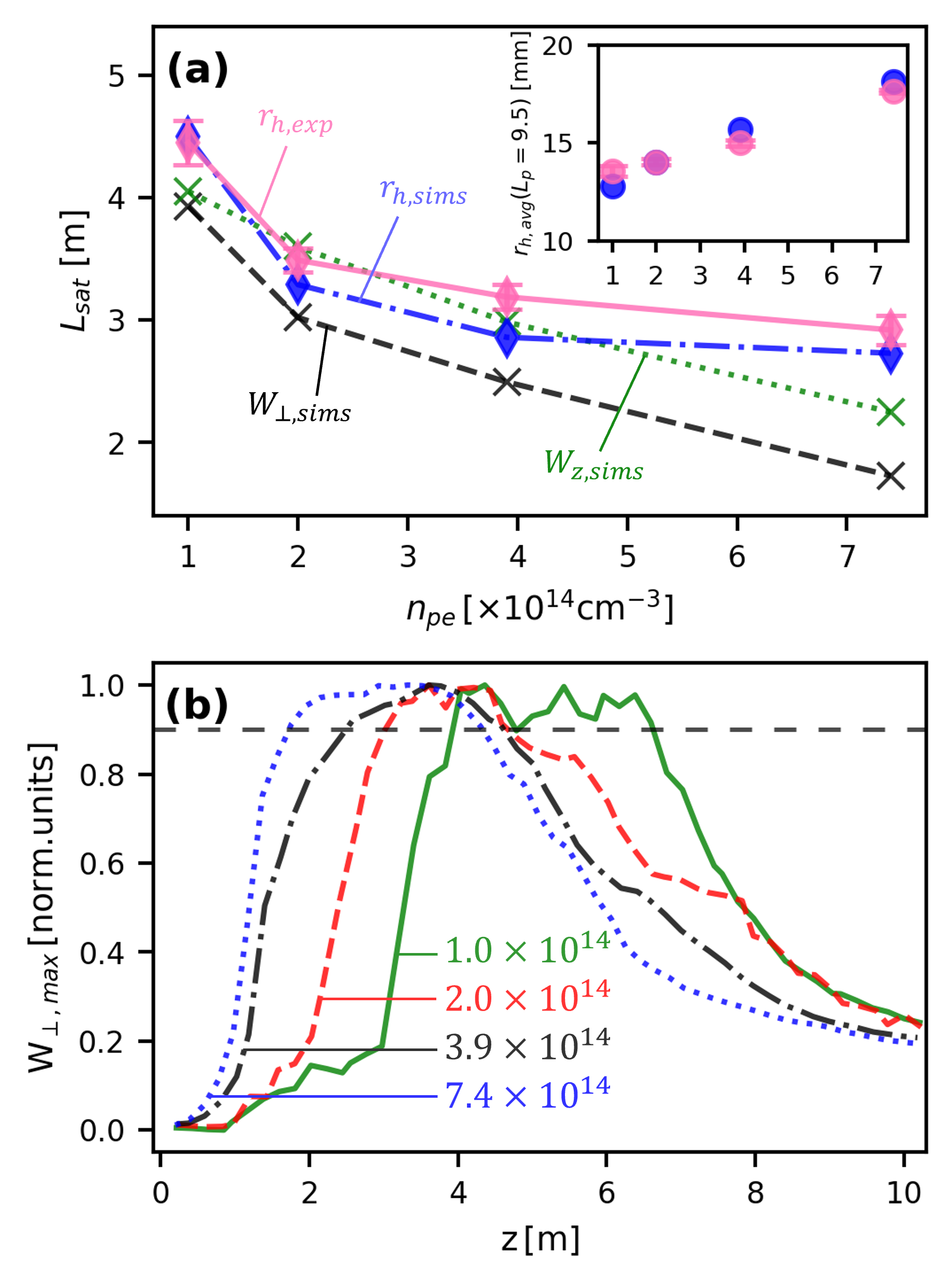}
    \caption{(a) Saturation length $L_{sat}$ as a function of $n_{pe}$ (same values as on Fig.~\ref{fig:waterfall}) labeled with the quantity it is determined from. %
    Inset: $r_{h,\mathrm{avg}}(L_p=9.5\,\text{m})$ as a function of $n_{pe} \, [\times 10^{14}\,$cm$^{-3}]$. %
    (b) Normalized transverse wakefield amplitude $W_{\perp,\mathrm{max}}$ along the plasma for the four $n_{pe}$ values of Fig.~\ref{fig:waterfall}. %
    Black dashed line: $90\%$ of peak value. %
    }
    \label{fig:saturation_length}
\end{figure}

To determine whether the saturation length of the halo radius provides an estimate for that of SM, we perform 2D axisymmetric particle-in-cell simulations using LCODE~\cite{LCODE} with parameters similar to those of the experiments.
We propagate ballistically the bunch phase space distribution at the end of the plasma $L_p$ to a location equivalent to that of the screen in the experiment. %
This produces transverse proton distributions that can be compared to those obtained in the experiment, e.g., Figs.~\ref{fig:experimental_setup}(a) and (b).
We determine the halo radius and the saturation length from these propagated distributions, using the same procedure as for the experimental images. %

The inset of Fig.~\ref{fig:saturation_length}(a) shows close agreement between experimental (pink symbols) and numerical simulation (blue symbols) results for $r_{h,\mathrm{avg}}(L_p=9.5\,\text{m})$, with differences below $1\,$mm for radii of more than $12\,$mm. %
The figure shows that the corresponding experimental and simulation saturation lengths of $r_h$ are also in very good agreement, with differences of less than $0.5$\,m. %
These small differences may be due to differences in the exact input parameters between numerical simulations and experiments~\cite{mariana}, namely bunch transverse size and emittance~\cite{gorn}. %

Numerical simulation results also provide the amplitude of the transverse wakefields $W_\perp$ that lead to the formation of the halo. %
Their amplitude varies along the bunch, and we plot on Fig.~\ref{fig:saturation_length}(b) their maximum amplitude as a function of position along the plasma, evaluated at $r=c/\omega_{pe}$ (the plasma electron collisionless skin depth with $\omega_{pe}=2\pi/\tau_{pe}$), and in the front of the bunch ($-\sigma_t <t < t_\mathrm{RIF}$, where $t=0$ is the bunch center). %
Simulation results show that this temporal extent of the bunch corresponds to the origin of the most defocused protons, i.e., those that define the halo radius.
In the context of a plasma wakefield accelerator, this extent also corresponds to the location in the wakefields along the bunch where a witness bunch would be injected for acceleration~\cite{lotovuniformity}. %
The field amplitudes are normalized to their maximum value to facilitate comparison of their development for different values of $n_{pe}$. %
In all cases, the wakefields grow, saturate, and then decay, which is typical of the development of SM along a plasma with constant density~\cite{kumar,lotov2015physics}.
It is clear from Fig.~\ref{fig:saturation_length}(b) that the development of SM occurs earlier with larger $n_{pe}$, as expected~\cite{schroeder}.
For consistency, we define the saturation length of the wakefields (and thus of SM) as the location where their amplitude reaches $90\%$ of their maximum value. %
We plot the saturation length of the wakefields on the same plot as that of the halo radius (black symbols on Fig.~\ref{fig:saturation_length}(a)). %
Although not the cause of the bunch halo, we apply the same analysis procedure to the longitudinal wakefields $W_z=E_z$ ($r=0$, and accelerating for electrons), and also plot their saturation length on Fig.~\ref{fig:saturation_length}(a) (green symbols). %

Figure~\ref{fig:saturation_length}(a) shows that the three saturation lengths follow the same trend versus $n_{pe}$ and remain close in value, although $W_\perp$ saturate earlier. %
The difference between $L_\mathrm{sat}$ inferred from $r_h$ and from $W_\perp$ may be due to the fact that, after saturation, the continuous backward shift of the wakefields, with phase velocity slower than that of the protons~\cite{pukhov,schroeder}, leads to further evolution of the SM process (see Fig.~\ref{fig:saturation_length}(b)).
As a result, protons from the microbunch train may enter the defocusing regions of the wakefields, gain additional transverse momentum, and exit the wakefields transversely. %
Because these defocused protons experience the largest field amplitudes, i.e., those near saturation, they reach the largest radii at the screen. %
This causes the halo radius to saturate slightly later than $W_\perp$, while still tracking its saturation length. %
Further along the plasma, the wakefield amplitude drops significantly, and protons shed by the backward shift of the wakefields do not acquire enough momentum to reach larger radii~\cite{NIMAarthur}. %

Overall, the results of Fig.~\ref{fig:saturation_length}(a) show that measuring the saturation length of the halo radius gives a good estimate of the saturation length of SM.
This method can therefore be used to determine $L_\mathrm{sat}$, a fundamental parameter of SM. %

\begin{figure}[b!]
    \centering
    \includegraphics[width=0.97\linewidth]{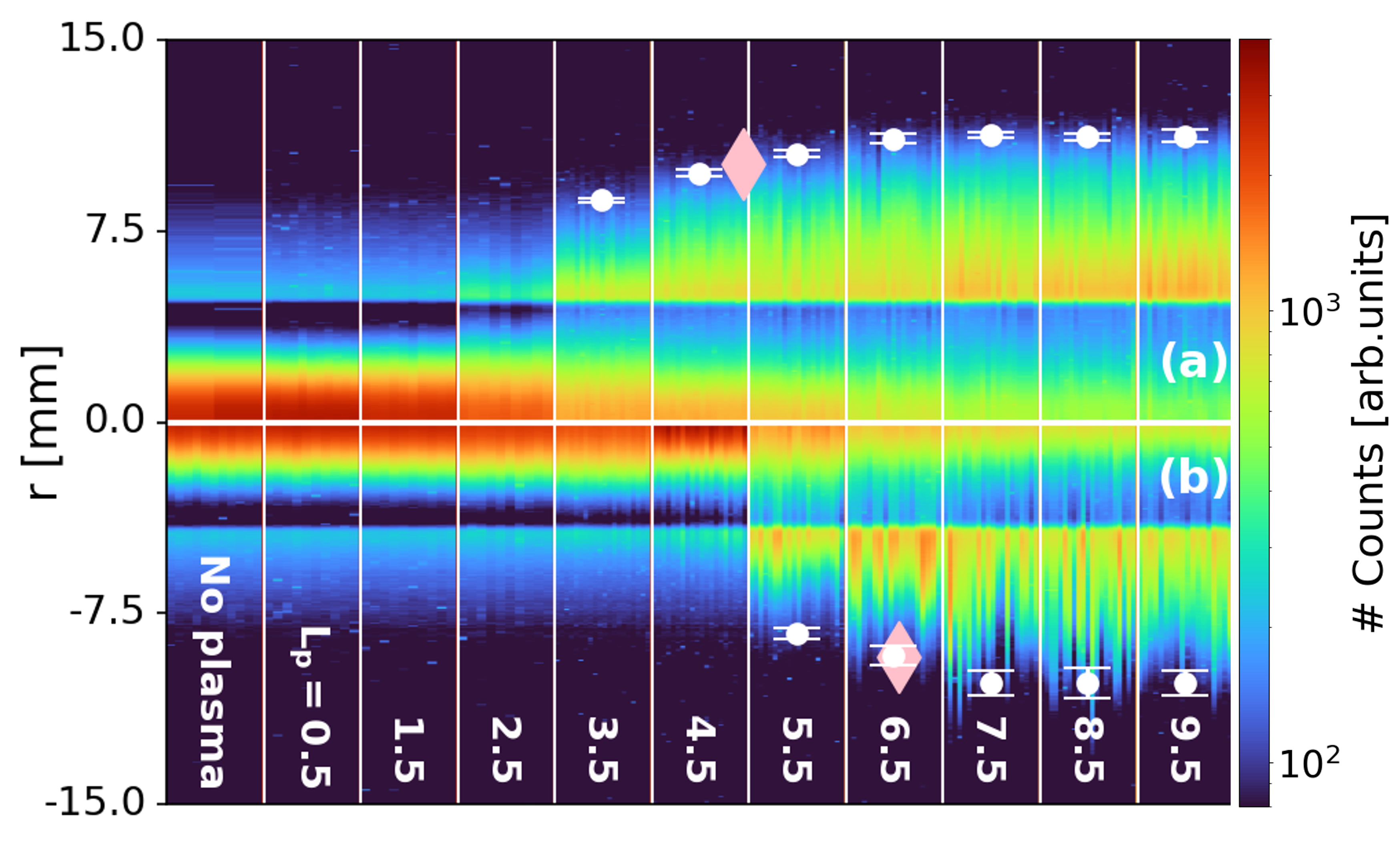}
    \caption{Radial slices of the transverse distribution of the bunch at the screen for different propagation distance in plasma ($L_p$). %
    (a) SSM ($t_\mathrm{RIF}=350\,$ps ($\sim2.1\sigma_t$)),
    (b) SMI ($t_\mathrm{RIF}=550\,$ps ($\sim3.2\sigma_t$)).
    Colormap: logarithmic scale. %
    White symbols: average radius of the halo $r_h$ for the corresponding $L_p$, error bars: standard deviation of $\sim20$ events. %
    Pink diamonds: saturation point of the radius ($r_h = 0.9 \,r_{h,\mathrm{max}}$). %
    (a) $n_{pe} = (1.02 \pm 0.01)\times10^{14}\,$cm$^{-3}$
    , $N_b = (2.92 \pm 0.02)\times10^{11}$, 
    (b) $n_{pe} = (1.08 \pm 0.01)\times10^{14}\,$cm$^{-3}$
    , $N_b = (2.87 \pm 0.02)\times10^{11}$.
    }
    \label{fig:SSMSMI_sat_length}
\end{figure}

We next investigate the effect of the initial wakefield amplitude on the saturation length of SM.
Since the initial wakefield amplitude scales with the local bunch density $n_b$ at the RIF~\cite{keinigs1987two,fabian}, it can be varied by changing the timing of the RIF along the bunch. %

Here, we compare the evolution of SM for $t_\mathrm{RIF}=350\,$ps (Fig.~\ref{fig:SSMSMI_sat_length}(a)) and $t_\mathrm{RIF}=550\,$ps (Fig.~\ref{fig:SSMSMI_sat_length}(b)). %
These $t_\mathrm{RIF}$ values are chosen such that the bunch charge in plasma (i.e., after the RIF) remains similar in the two cases (difference $<2.5\,\%$), ensuring that the dominant varying parameter is the initial field amplitude, and not the growth rate~\cite{kumar,pukhov,schroeder}. %
For $t_\mathrm{RIF}=550\,$ps, the initial field amplitude induced by the RIF is lower than for $t_\mathrm{RIF}=350\,$ps ($W_{\perp,0}(t_\mathrm{RIF})\propto n_b(t_\mathrm{RIF})$), and not sufficient to overcome the wakefields driven by noise or features in the bunch distribution. %
In this case, the SM process is no longer seeded and develops as an instability (SMI)~\cite{fabian,lotovnoise}. %

In the seeded self-modulation (SSM) case, the wakefields acquire defocusing regions and reach a high amplitude earlier along $L_p$, as shown on Fig.~\ref{fig:SSMSMI_sat_length} by the halo of defocused protons appearing earlier along the plasma ($L_p \approx 3.5\,$m, Fig.~\ref{fig:SSMSMI_sat_length}(a)) than in the SMI case ($L_p \approx 5.5\,$m, Fig.~\ref{fig:SSMSMI_sat_length}(b)).
Although the variations in initial field amplitude and asymmetries in the modulated bunch structure are larger in the case of SMI (visible on the images, not shown here), the radius of the halo also grows and saturates, as shown by the average radius values (white symbols in Fig.~\ref{fig:SSMSMI_sat_length}(b)).
Following the same procedure as the one used on Fig.~\ref{fig:saturation_length}(a), we determine that the halo radius saturates at $L_\mathrm{sat} = (4.9\pm0.2)\,$m in the SSM case and $L_\mathrm{sat} = (6.6\pm0.4)\,$m in the SMI case, as indicated on Fig.~\ref{fig:SSMSMI_sat_length} with pink diamonds.
This result shows that seeding, i.e., providing higher initial field amplitude, reduces the saturation length of SM. %
This dependency is further supported by comparing $t_\mathrm{RIF}=100\,$ps (Fig.~\ref{fig:waterfall}(a)) and $t_\mathrm{RIF}=350\,$ps (Fig.~\ref{fig:SSMSMI_sat_length}(a)), where saturation occurs earlier at $t_\mathrm{RIF}=100\,$ps ($L_\mathrm{sat} = (4.5\pm0.2)\,$m vs $(4.9\pm0.2)\,$m), i.e., when initial wakefields are larger, despite the lower bunch charge in plasma. %

Figure~\ref{fig:SSMSMI_sat_length} further shows that $r_h$ is more reproducible from event to event in the SSM case than in the SMI case, e.g., with variations of $2.0$\% versus $4.6$\% (one standard deviation, shown with the error bars) for $L_p=9.5\,$m. %
This is consistent with timing reproducibility measurements of the microbunch train shown in the seeding demonstration experiments~\cite{fabian,livio}, and will be the subject of another publication.

\emph{Summary \& Conclusions} -- We have experimentally determined the saturation length of the self-modulation instability by measuring the evolution of the radius of the halo of defocused protons versus plasma length. %
We used numerical simulation results to show that saturation of the halo radius also corresponds to saturation of longitudinal and transverse wakefields, all three with similar saturation lengths. %
The saturation length decreases with increasing plasma electron density and initial field amplitude, e.g., when seeding. %
Understanding and measuring the characteristics of instabilities is key for validating theoretical models and numerical simulation results. %
For plasma-based accelerators driven by a long particle bunch using SM, measuring this length is crucial, as external injection must occur after saturation to maximize witness bunch energy and quality~\cite{pukhov}. %
In particular, we have shown that SM saturates well within the $10\,$m of plasma foreseen for the length of the self-modulator in future acceleration experiments in AWAKE, which aims to produce high-energy electrons for particle physics applications~\cite{CaldwellLowLumi,Caldwell2016,Wing2019}. %
Measuring the saturation length is also very important for plasma-based accelerators driven by long laser pulses developed as X-ray and $\gamma$-ray radiation sources~\cite{albert,SMLWFAexp2}.%

\begin{acknowledgments}
    This work was supported in parts by Fundação para a Ciência e Tecnologia–Portugal (Grant No. CERN/FIS TEC/0017/2019, No. CERN/FIS-TEC/0034/2021, No. UIBD/50021/2020); STFC (AWAKE-UK, Cockcroft Institute core, John Adams Institute core, and UCL consolidated grants), United Kingdom; the National Research Foundation of Korea (Grant No. NRF-2016R1A5A1013277 and No. NRF-2020R1A2C1010835). M. W. acknowledges the support of DESY, Hamburg. Support of the Wigner Datacenter Cloud facility through the Awakelaser project is acknowledged. UW Madison acknowledges support by NSF Award No. PHY-1903316. The AWAKE collaboration acknowledges the SPS team for their excellent proton delivery.
\end{acknowledgments}

\bibliographystyle{unsrt}
\bibliography{refs.bib}

\end{document}